\documentstyle[preprint,aps,prd,psfig]{revtex}
\begin{document}

\title{Chemical Evolution of Strongly Magnetized Quark Core\\
in a Newborn Neutron Star}
\author{Tanusri Ghosh$^{a)}$\thanks{E-mail:tanusri@klyuniv.ernet.in}
and Somenath Chakrabarty$^{a,b)}$\thanks{E-mail:somenath@klyuniv.ernet.in}\\
 a) Department of Physics, University of Kalyani, West Bengal 741 235,
India\\
b) Inter-University Centre for Astronomy and Astrophysics, Post Bag 4\\
Ganeshkhind, Pune 411 007, India}

\date{\today}
\maketitle
\begin{flushleft}
\end{flushleft}

\begin{abstract}

The chemical evolution of nascent quark matter core in a newborn compact
neutron star is studied in presence of a strong magnetic field. The
effective rate of strange quark production in degenerate quark matter core in
presence of strong magnetic fields is obtained. The investigations show that
in presence of strong magnetic fields a quark matter core becomes  
energetically unstable and hence a deconfinement transition to quark matter 
at the centre of a compact neutron star under such circumstances is not 
possible. The critical strength of magnetic field at the central core to make
the system energetically unstable with respect to dense nuclear matter
is found to be $\sim 4.4\times 10^{13}$G.  This is the typical strength
at which the Landau levels for electrons are populated. The other
possible phase transitions at such high density and ultra strong
magnetic field environment are discussed.
\end{abstract}

\newpage
\section{Introduction}
More than a decade ago, it was suggested by Witten \cite{wi1} that a flavor
symmetric mixture of $u$, $d$ and $s$-quarks known as strange quark
matter (SQM) would be the most stable configuration of hadronic matter. An
extensive theoretical investigations on the stability of both bulk as well as
quasi-bulk properties of SQM using MIT bag model and also with some other
confinement models have proved the reality of such speculation
\cite{alc1,han1,sc0}. Usually  the strange quark is attributed a minor role in
the standard model of the Universe. However, it has also been suggested by
Witten that strange quark matter might be an important relic from the
early Universe and also a constituent of neutron stars. Although strange quark
matter could be absolutely stable and thus be the true ground state of
hadronic matter, long-lived or stable strange quarks are only found in the
regions with very high density, otherwise, the relatively large $s$-quark
mass and the weak mixing to the down quarks lead to
rapid decay. In a quark matter system, either at the core of a compact
neutron star or in the early Universe-microsecond after big bang,
$s$-quarks are produced through a number of weak processes. It is generally
believed that microsecond after big bang, the Universe was
completely filled with a hot and dense soup of quark matter. The
survival of strange quark nuggets relic till today from the early Universe
is a controversial issue \cite{mad1}. It is also expected that a
deconfinement transition to bulk quark matter may cur at the core of
a neutron star if the density exceeds a few times normal nuclear matter
density. In such a cold quark matter system, $s$-quarks are produced
through weak processes. Although the existence of strange quark
nuggets relic produced as a result of first order QCD phase transition in
the early Universe is highly uncertain, the
presence of exotic quark matter core in a compact neutron star can not
be ruled out. A few years ago, Madsen and Hiselberg et. al. have investigated 
with detail numerical calculations the rate of $s$-quark production in dense
quark matter in the context of survival of strange quark nuggets produced at 
the time of primordial QCD phase transition 
\cite{mad1,mad2}. Recently, Glendenning and his group have reported a lot of
exciting results on the existence of strange quark matter core in a
compact neutron star and its effect on various gross-properties of hybrid
stars \cite{gln1}. It has also been suggested that a transition to quark
matter core may occur during slowing down of a old neutron star \cite{hen}.
A first order phase transition to quark matter at the core of a neutron
star is also possible by seeding strange droplets from outside and by
neutrino sparking \cite{horv}.  However, in
none of these papers, the effect of strong magnetic field of the neutron
star (particularly in the case of newborn neutron star) on such exotic
matter have been considered. Large magnetic fields are known to be
present in neutron stars \cite{chanm,bar1,kou1}. 
The direct evidence from pulsar observations
show that the strength of this magnetic field at the neutron star
surface is $10^{12}-10^{14}$G. Then it can very easily be shown by the scalar
virial theorem that magnetic field strength at the core region of a
neutron star may reach up to $10^{18}-10^{19}$G \cite{lai1}. On the other hand,
from the stability criteria of a compact magnetized neutron star it can very
easily be shown following the famous work by Chandrasekhar and Fermi 
\cite{chan} in the context of magnetized white dwarfs, that the upper
limit for the strength of magnetic field at the core region is $\sim
10^{19}$G. The presence of such strong magnetic field inside newborn neutron
stars significantly modifies the order of qaurk-hadron phase
transition at the core. It has been shown that a first order quark-hadron phase
transition initiated by the nucleation (by thermal or quantum
fluctuation) of quark droplets in a system of meta-stable neutron matter
in presence of a strong magnetic field
($\geq B_m^{(c)(e)}$) is absolutely forbidden \cite{sc1,sc2}. However, 
some higher
order transitions, e.g. metal-insulator type of second order phase transition
is allowed if the strength of magnetic field is $\leq 10^{20}$G. Which
is of course too high to achieve at the core of a neutron star. Even if
the transition is of second order in nature, the gross properties of
dense quark matter are considerably modified with respect to the field
free case. The presence of strong magnetic field also affects the weak
and electro-magnetic processes which may occur in the degenerate quark
matter core \cite{sc3}. The presence of strong magnetic field shifts the
$\beta$-equilibrium point of the quark matter system \cite{sc4}. It also affects
the abundances of various components in $\beta$-equilibrium.
The photon emission by bremsstrahlung, photon splitting and
neutrino emission processes are significantly affected by the presence of
strong magnetic field \cite{sc3,hey}.
It is also believed that pulsar kicks are
generated by the asymmetric  neutrino emission in presence of strong
magnetic fields \cite{vil,horw,lai2,raf}. In a dense quark matter, since 
$s$-quarks are produced or annihilated through weak processes, the presence of 
strong magnetic field should also alter the production and decay rates of
$s$-quarks. To the best of our knowledge, theoretical investigation of this
particular problem has not been reported
earlier. In the previous work by Madsen and Hiselberg et. al. \cite{mad1,mad2}
the rate of $s$-quark production in degenerate cold quark matter ($T<10$MeV) was
studied for the field free case in the context of stability of strange nuggets as
the relic of big bang deconfinement transition. In the present paper we shall
study the effect of strong magnetic field on the net rates of $u$, $d$ and
$s$-quarks production and there by study the chemical evolution of nascent 
degenerate quark matter core in a compact neutron star. We shall also consider 
the effect of neutrino trapping in degenerate quark matter on the chemical 
evolution in presence of a strong magnetic field. The main finding of
this paper is that the chemical evolution of nascent quark matter core
in presence of a strong magnetic field leads to an unphysical scenario-
the final state in $\beta$-equilibrium is energetically unstable with 
respect to the normal hadronic matter of neutron star.
Although a flavor symmetric mixture of $u$, $d$ and $s$-quarks at zero
temperature is energetically much more stable than the corresponding
field free case, the chemical evolution of a bulk system consisting of
$u$, $d$ and $s$-quarks (not a flavor symmetric mixture) shows that the
quark matter core of a newborn neutron star may become energetically
unstable if the magnetic field strength exceeds a typical value $\sim
4.4\times 10^{13}$G, which is the strength of magnetic field at which
the Landau levels for electrons are populated and quantum mechanical
effect of strong magnetic field on electrons become important. Since the
formation of a flavor symmetric quark matter at the central region of a
compact neutron star is possible only through the chemical evolution of
ordinary quark matter or $u$, $d$ and $s$ flavor asymmetric quark
matter, it would really be interesting to investigate the time evolution
of various species in the system both for field free case and in
presence of a strong magnetic field.\par
For the sake of simplicity we   neglect quark-quark interaction in the
degenerate quark matter. The temperature of the system is taken to be 
$1$MeV $(\simeq 10^9K)$. The increase or decrease in temperature of the
system causes an enhancement or suppression of effective production rate
of $s$-quark respectively. However, it only makes a quantitative change in our
result. The qualitative nature remains unaltered.  It is further assumed that 
the external magnetic field $B_m$ is constant throughout the matter.
The convenient choice of gauge is $A_0=0$ and
$\vec A=(0,xB_m,0)$ corresponding to the constant magnetic field $B_m$
along z-axis. If the strength of magnetic field $B_m$ is greater than
the critical value $B_m^{(c)(i)}=m_i^2/q_i$ for the $i$-th charged
particle, where $m_i$ and $q_i$ are respectively the mass and charge of
the $i$-th species, the corresponding single particle energy is given by
$\varepsilon _i=(p_z^2 +m_i^2 +2q_i\nu B_m)^{1/2}$, otherwise the usual
expression for field free $(B_m=0)$ case is used, where $\nu$ is the
Landau quantum number. The critical strength of magnetic
field as mentioned above can be
obtained by equating the cyclotron quantum with the rest mass energy of the
$i$-th charged particle. If the magnetic field strength exceeds this typical
value, the quantum mechanical effect of strong magnetic field on the
$i$-th charged component becomes important. This is known as Landau
diamagnetism. For electrons of mass $0.5$MeV, this typical strength is
$B_m^{(c)(e)}\simeq 4.4\times 10^{13}$G, whereas for $u$ and $d$-quarks
of current mass $5$MeV, it is $\simeq 10^2 B_m^{(c)(e)}$. On the other
hand for $s$-quark of current mass $150$MeV, this typical value is $\sim
10^{20}$G. Which is too high to achieve even at the core of a
newborn neutron star. Therefore in the case of $s$-quarks, quantum
mechanical effect of strong magnetic field is insignificant and results for
the field free case are used.\par
The paper is organized in the following manner, in section 2 we have given
the basic formalism to study the chemical evolution of compact and degenerate
nascent quark matter core with non-degenerate neutrinos (i.e. neutrinos are
assumed to leave the system freely immediately after their formation), 
both in absence and presence of strong magnetic fields. In the case of
$B_m\neq 0$, we have considered two possible scenarios-(i) magnetic field
strength is low enough so that
only  electrons are affected and (ii) the field strength is relatively
stronger so that all the charged particle components except $s$-quarks are
affected. In section 3 we have discussed with detail numerical calculations,
the effect of neutrino trapping on chemical evolution of the system for
both $B_m=0$ and $B_m \neq 0$ cases.
Concluding remarks are presented in section 4.
\par
\section{Basic Formalism: With non-degenerate Neutrinos}
To study the chemical evolution of degenerate quark matter at the core
of a newborn neutron star we consider the following weak interaction processes
which are taking place at the quark matter core
\begin{equation}
d\rightarrow u+e^- +\bar \nu_e
\end{equation}
\begin{equation}
u+e^- \rightarrow d+\nu_e
\end{equation}
\begin{equation}
s\rightarrow u+e^- +\bar \nu_e
\end{equation}
\begin{equation}
u+e^- \rightarrow s+\nu_e
\end{equation}
\begin{equation}
u+d\rightleftharpoons u+s
\end {equation}
We further assume that the system is charge neutral and produced
neutrinos/anti-neutrinos leave the system freely (i.e. they are
non-degenerate). The approach to chemical equilibrium of the system is
governed by the following sets of kinetic equations
\begin{equation}
\frac{dY_u}{dt}=\frac{1}{n_B}[\Gamma_1-\Gamma_2 +\Gamma_3 -\Gamma_4]
\end{equation}
\begin{equation}
\frac{dY_d}{dt}=\frac{1}{n_B}[-\Gamma_1 +\Gamma_2 -\Gamma_5 +\Gamma_6]
\end{equation}
Baryon number conservation and charge neutrality conditions relate strange 
quark and electron abundances with that of $u$ and $d$-quark and are
given by
\begin{equation}
Y_s =3-Y_u -Y_d
\end{equation}
\begin{equation}
Y_e =Y_u -1
\end {equation}
where $Y_i=n_i /n_B $ is the abundance for the $i$-th component ($i=u,
d, s, e$ and $\nu_e$ (for the degenerate neutrino case only)), $n_B$ is
the baryon number density, $\Gamma_i$ is the rate of $i$-th weak
interaction process ($i=1,...,5$) and $\Gamma_6$ is the rate of reverse
process as indicated by eqn.(5).\par
To study chemical evolution of the system we solve the
differential eqns.(6) and (7) numerically along with two
constraint equations, eqns.(8) and (9) and use the rates $\Gamma_1$
to $\Gamma_6$, appear on the right hand side of eqns.(6) and (7). Since
these differential equations are of first order in nature, we  need
at least one initial condition for both $Y_u$ and $Y_d$ in order to get
the complete solutions. For a neutron star of mass $\simeq 1.4M_{\odot}$ the
baryon number density $n_B$ at the centre is of the order of $3-4 n_0 $,
temperature $\sim 10^9$K and proton content is about $4\%$. Then the initial
conditions are given by $Y_u(t=0)=1.04, Y_d(t=0)=1.96$. As a consequence of
baryon number and charge conservation we have  $Y_s(t=0)=0$ and
$Y_e=0.04$.
\subsection{Weak Reaction Rates: $B_m=0$}
In this subsection we shall first derive the rates ($\Gamma_1$ to $\Gamma_5$)
for the processes given by
eqns.(1)-(5) and also the rate $\Gamma_6$ for the reverse process as
indicated  by eqn.(5), for the field free case. \par
The reaction rate for the process (1) is given by
\begin{eqnarray}
\Gamma_1 &=& 6\times 64G^2\cos^2 \theta_c \int\sum_{i=1}^4
\frac{d^3p_i}{(2\pi)^3 2\varepsilon_i} (2\pi)^4\delta^4(p_1 -p_2 -p_3 -p_4)
\nonumber \\
& & f(\varepsilon_1)(1-f(\varepsilon_3))(1-f(\varepsilon_4))(p_1
.p_2)(p_3 .p_4)
\end{eqnarray}
where $i=1, 2, 3, 4$ stand for $d$, $\bar\nu_e$, $u$ and $e$
respectively, $f(\varepsilon_i)=1/(1+\exp({(\varepsilon_i
-\mu_i)/T)}$, the equilibrium (Fermi) distribution function for the $i$-th
component and $p_i$, $\varepsilon_i$, $\mu_i$ are respectively the 
$4$-momentum, energy and chemical potential of the particle $i$, G is the weak
coupling constant, $\theta_c$ is the Cabibbo angle $(\cos^2
\theta_c \simeq 0.74)$ and T is the temperature of the system. The
angular part of the integrals can be evaluated without any approximation
\cite{mad1,mad2} and we get
\begin{eqnarray}
\Gamma_1 &=& \frac{12G^2}{\pi^6}\cos^2 \theta_c \int \sum_{i=1}^4 d\varepsilon_i
\delta(\varepsilon_1 -\varepsilon_2 -\varepsilon_3 -\varepsilon_4)
\nonumber \\
& & f(\varepsilon_1)(1-f(\varepsilon_3))(1-f(\varepsilon_4))
I_n(\varepsilon_1, \varepsilon_2, \varepsilon_3, \varepsilon_4)
\end{eqnarray}
where
\begin{eqnarray}
I_n (\varepsilon_1, \varepsilon_2, \varepsilon_3, \varepsilon_4) =
k_- ^{12}k_+ ^{34} I_1(\varepsilon_1, \varepsilon_2, \varepsilon_3,
\varepsilon_4) +
 &&\frac{1}{6}(k_+ ^{34}-k_- ^{12})I_2(\varepsilon_1, \varepsilon_2,
\varepsilon_3, \varepsilon_4)-\nonumber \\
&&\frac{1}{20} I_3(\varepsilon_1,
\varepsilon_2, \varepsilon_3, \varepsilon_4)
\end{eqnarray}
\begin{equation}
k_{\pm}^{ij}=\varepsilon_i \varepsilon_j
\pm \frac{1}{2} (p_i^2 +p_j^2)
\end{equation}
\begin{eqnarray}
I_n(\varepsilon_i, \varepsilon_j, \varepsilon_k, \varepsilon_l) &=&
(P_{ij}^{2n-1}-p_{ij}^{2n-1})\theta(P_{kl}-P_{ij}) \theta(p_{ij}-p_{kl})
\nonumber \\
& & +(P_{ij}^{2n-1}-p_{kl}^{2n-1})\theta(P_{kl}-P_{ij})
\theta(p_{kl}-p_{ij})\theta(P_{ij}-p_{kl}) \nonumber \\
& & +(P_{kl}^{2n-1}-p_{kl}^{2n-1})\theta(P_{ij}-P_{kl})
\theta(p_{kl}-p_{ij})\nonumber \\
& & +(P_{kl}^{2n-1}-p_{ij}^{2n-1})\theta(P_{ij}-P_{kl})
\theta(p_{ij}-p_{kl}) \theta(P_{kl}-p_{ij})
\end{eqnarray}
$P_{ij}=p_i +p_j$, $p_{ij}=\mid p_i -p_j \mid$, $p_i =\mid \vec p_i
\mid$ is the magnitude of three momentum vector and
$\varepsilon_i=(p_i^2 +m_i^2)^{1/2}$ (we assume $\hbar=c=k_B =1$).\par
In the case of a compact neutron star, the central density is very high 
($\sim 3-4n_0$ to have
a deconfinement transition) but the temperature is low enough ($\sim 10^9$K),
as a consequence $u$, $d$ quarks and electrons are relativistic in nature
and are strongly
degenerate. Under such an extreme condition all the integrals except one
in eqn.(11) can be evaluated analytically and the final expression for rate
of the process (1) is given by
\begin{equation}
\Gamma_1(\mu_d ,\mu_u ,\mu_e ,T)=\frac{3G^2}{4\pi^5}\cos^2 \theta_c
\int_{0}^{\infty}dx \frac{(x-\xi_d)^2 +\pi^2}{1+\exp{(x-\xi_d)}}
I(\mu_d, Tx, \mu_u ,\mu_e)
\end{equation}
where in the expression for $I$ we have replaced energy and momentum of
all the particles except the neutrino by their values on the Fermi
surface, i.e. we put $\varepsilon_i=\mu_i$ and $p_i=p_{F_i}=(\mu_i^2
-m_i^2)^{1/2}$, we have also changed the integration variable
$\varepsilon_{\nu}$ to $x$, where neutrino energy 
$\varepsilon_{\nu}=p_{\nu}=xT$ and $\xi_d=(\mu_d -\mu_u -\mu_e)/T$ which
is zero in the $\beta$-equilibrium condition. Following the same
methodology, we have obtained the rate for the process (2), given by
\begin{equation}
\Gamma_2(\mu_d ,\mu_u ,\mu_e ,T)=\frac{3G^2}{4\pi^5}\cos^2\theta_c T^3
\int_{0}^{\infty}dx \frac{(x+\xi_d)^2 +\pi^2}{1+\exp{(x+\xi_d)}} J(\mu_d
, Tx, \mu_u ,\mu_e)
\end{equation}
where
\begin{eqnarray}
J(\varepsilon_1, \varepsilon_2, \varepsilon_3, \varepsilon_4) &=&
k_+^{12}K_+^{34} I_1(\varepsilon_1, \varepsilon_2, \varepsilon_3,
\varepsilon_4) \nonumber \\
& & -\frac{1}{6}(k_+^{34}+k_+^{12}) I_2(\varepsilon_1, \varepsilon_2,
\varepsilon_3, \varepsilon_4)+\frac{1}{20}I_3(\varepsilon_1,
\varepsilon_2, \varepsilon_3, \varepsilon_4)
\end{eqnarray}
The reaction rates for the processes (3) and (4) corresponding to
$s$-quark can very easily be evaluated by replacing all the parameters
for $d$-quarks by the corresponding $s$-quark values (i.e.
$\mu_d \rightarrow\mu_s$, $\xi_d\rightarrow\xi_s$, $m_d
\rightarrow m_s$) and $\cos^2 \theta_c$ by $\sin^2\theta_c$ in the
expressions for $\Gamma_1$ and $\Gamma_2$ respectively.\par
For the weak process (5) we have
\begin{eqnarray}
\Gamma_5(\mu_d, \mu_u, \mu_s, T) &=& \frac{9G^2}{2\pi^5} \cos^2\theta_c
\sin^2 \theta_c T^3 \int_{-\infty}^{(\mu_s -m_s)/T}dx
\nonumber \\
& & \frac{(x+\xi_s)^2+\pi^2}{(1+\exp (x))(1+\exp(-x_x -\xi_s))} J(\mu_d,
\mu_u, \mu_u, \mu_s -Tx)
\end{eqnarray}
where $\xi_s =(\mu_d -\mu_s)/T$.\par
The reaction rate for the reverse process as shown in eqn.(5) is related to
the forward rate $\Gamma_5$ by
\begin{equation}
\Gamma_6=\exp{(-\xi_s)}\Gamma_5
\end{equation}
\subsection{Weak Reaction Rates: $B_m\neq 0$}
We shall now evaluate the rates $\Gamma_1$ to $\Gamma_6$ in presence of 
strong magnetic fields. 
We consider two possible cases: the magnetic field $B_m$ is not very
high ($B_m^{(c)(e)} < B_m <B_m^{(c)(u,d)}$)
so that the quantum mechanical effect is important for electrons only,
and finally the magnetic field is strong enough to affect all the
charged particle components except $s$- quarks ($B_m^{(c)(u,d)} <B_m
<B_m^{(c)(s)}$) .\par
Let us first consider case 1. The magnetic field
$B_m^{(c)(e)}<B_m<B_m^{(c)(u,d)}$. To evaluate the rates for the weak
processes (1)-(4), we use the modified form of spinor solution of Dirac
equation in presence of a strong magnetic field for
electron \cite{sc2,sc5,sc6}. For other charged particles, the usual 
plane wave Dirac spinor solutions for field free case are used. Since only
electrons are affected, the rates for 
the process (5) and its reverse process will remain unchanged.\par
In this case the rate of the process (1) is given by
\begin{eqnarray}
\Gamma_1 &=& \frac{3G^2 q_e B_m}{2\pi^6}T^4 \cos^2\theta_c
\sum_{\nu=0}^{\infty} \mu_u \mu_e \left(\frac{p_{F_u}}{p_{F_e}}\right)
\nonumber\\
& & \int_{-\infty}^{+\infty}\int_{-\infty}^{+\infty}\left(x_u +x_e
-\frac{\mu_u +\mu_e -\mu_d}{T}\right)^2 f(x_u)f(x_e) dx_u dx_e
\end{eqnarray}
where $q_e$ is the charge carried by electron,
$f(x_i)=1/(1+\exp(x_i))$, and $p_{F_e}=(\mu_e^2 -m_e^2 -2\nu q_e
B_m)^{1/2}$ is the Fermi momentum for electron. The sum $\nu$ is over
Landau quantum number for the electron which in principle can have all 
possible positive integer values including zero.\par
Similarly the rate for the process (2) is given by
\begin{eqnarray}
\Gamma_2 &=& \frac{3G^2}{2\pi^6}(q_e B_m)T^4 \cos^2\theta_c
\sum_{\nu=0}^{\infty}\mu_u
\mu_e \left(\frac{p_{F_u}}{p_{F_e}}\right)
\int_{-\infty}^{+\infty}\int_{-\infty}^{+\infty} dx_u dx_e \nonumber\\
& & \left(x_u +x_e +\frac{\mu_u +\mu_e -\mu_d}{T} \right)^2 f(x_u)
f(x_e)
\end{eqnarray}
As before the rates for the processes (3) and (4) may be obtained by
simply replacing
all the parameters for $d$-quarks by the corresponding $s$-quark values
and $\cos^2\theta_c$ by $\sin^2\theta_c$ in eqns.(20) and (21) respectively.
Whereas the rates for the direct and the reverse processes given by
eqn.(5) remain unaltered.\par
Next we consider the second case possibility, $B_m^{(c)(s)} > B_m >
B_m^{(c)(u,d)}$. In this case modified
spinor solutions for all the charged particle components except $s$-quark 
are used to evaluate rates of the weak processes \cite{sc2,sc5,sc6}. 
As we have noticed that compact analytical expressions 
for the rates can only be obtained if we assume that zeroth Landau levels are
populated for the magnetically affected charged species. Otherwise we
have to evaluate the rates numerically right from the beginning.\par
The rate for the process (1) is given by
\begin{eqnarray}
\Gamma_1 &=& \frac{G^2}{2\pi^5}(q_e B_m)\left[ \frac{2q_u B_m q_d
B_m}{5}\right]^{1/2} \cos^2\theta_c T^4 \frac{\mu_u
\mu_e}{p_{F_u}p_{F_e}} \nonumber\\
& & \int_{-\infty}^{+\infty}\int_{-\infty}^{+\infty}\left[x_u +x_e
+\frac{\mu_d -\mu_u -\mu_e}{T}\right]^2 f(x_u) f(x_e) dx_u dx_e
\end{eqnarray}
where $q_u$ and $q_d$ are respectively the charge carried by  $u$ and
$d$-quarks.\par
Similarly the rate for the process (2) is given by
\begin{eqnarray}
\Gamma_2 &=& \frac{G^2}{2\pi^5}(q_e B_m)\left[\frac{2q_u B_m q_d
B_m}{5}\right]^{1/2} \cos^2\theta_c T^4 \nonumber\\
& & \int_{-\infty}^{+\infty}\int_{-\infty}^{+\infty}\left[x_u +x_e
+\frac{\mu_u +\mu_e -\mu_d}{T}\right]^2 f(x_u) f(x_e) dx_u dx_e
\end{eqnarray}
In the present scenario, the rates for the processes (3) and (4) can not be 
obtained by simply replacing $d$-quark parameters with the corresponding 
$s$-quark values and $\cos^2\theta_c$ by $\sin^2\theta_c$ in eqns.(22) and (23)
respectively. Since $s$-quarks are not affected
by the magnetic field, we have to evaluate $\Gamma_3$ and
$\Gamma_4$ with zero field plane wave solution for $s$-quark and modified
Dirac spinors for all other charged species. Then we have the expressions
for rates of the processes given by eqns.(3) and (4) for zero Landau
quantum number
\begin{eqnarray}
\Gamma_3 &=& \frac{5G^2}{16\pi^7}(q_e B_m q_u B_m)T^4 \sin^2\theta_c
\int_{-\infty}^{+\infty}\int_{-\infty}^{+\infty}f(x_u) f(x_e)
\nonumber\\
& & \left(x_u +x_e +\frac{\mu_s -\mu_u -\mu_e}{T}\right)^2 dx_u dx_e
\end{eqnarray}
and
\begin{eqnarray}
\Gamma_4 &=& \frac{5G^2}{16\pi^7}(q_e B_m q_u B_m)T^4 \sin^2\theta_c
\int_{-\infty}^{+\infty}\int_{-\infty}^{+\infty}f(x_u) f(x_e)
\nonumber\\
& & \left(x_u +x_e -\frac{\mu_s -\mu_u -\mu_e}{T}\right)^2 dx_u dx_e
\end{eqnarray}
respectively.\par
The rate for the process (5) is given by
\begin{eqnarray}
\Gamma_5 &=& \frac{9G^2}{40\pi^6}(q_e B_m)^2 \cos^2\theta_c
\sin^2\theta_c \sqrt\frac{5}{6}~ \frac{\mu_u \mu_d}{p_{F_u}p_{F_d}}
\nonumber\\
& & T\left(1-\frac{p_{F_u}p_{F_d}}{\mu_u \mu_d}\right)\int_{x_u
=-\infty}^{+\infty} dx_u f(x_u) f(-x_u -x_s +\frac{\mu_d +\mu_s}{T})
\nonumber\\
& & \exp(-\frac{3}{5q_e B_m}p_{s_x}^2) \exp(-\frac{3}{2q_e
B_m}p_{s_y}^2)\left[1-\frac{(p_{F_u} +p_{F_d} -p_{s_z})p_{s_z}}{\mu_u
E_s}\right]f(x_s) d^3p_s
\end{eqnarray}
where $d^3p_s=p_s^2 dp_s \sin\theta d\theta d\phi$. Since
$E_s=\sqrt{p_s^2 +m_s^2}$, we have
\begin{equation}
d^3p_s=p_s E_s dE_s \sin\theta d\theta d\phi
\end{equation}
\begin{equation}
p_{s_x}=p_s \sin\theta \cos\phi
\end{equation}
\begin{equation}
p_{s_y}=p_s \sin\theta \sin\phi
\end{equation}
and
\begin{equation}
p_{s_z}=p_s \cos\theta
\end{equation}
To evaluate this integral we use $x_s =(\mu_s -E_s)/T$ as the new integration
variable instead of $E_3$. Then we have
\begin{eqnarray}
\Gamma_5 &=& \frac{9G^2}{40\pi^6}(q_u B_m)^2 \cos^2\theta_c
\sin^2\theta_c \sqrt\frac{5}{6}\frac{\mu_u \mu_d}{p_{F_u}p_{F_d}}
\nonumber\\
& & T^2 \left(1-\frac{p_{F_u}p_{F_d}}{\mu_u
\mu_d}\right)\int_{-\infty}^{+\infty}\int_{-\infty}^{+\infty}dx_u dx_s
\int_{0}^{\pi}\sin\theta d\theta \int_{0}^{2\pi}d\phi \nonumber\\
& & f(x_u)f(x_s)f\left(-x_u -x_s +\frac{\mu_d +\mu_s}{T}\right)
\left[1-\frac{(p_{F_u}+p_{F_d}-p_{s_z})p_{s_z}}{\mu_s E_s}\right]
\nonumber\\
& & \exp\left(-\frac{3}{5q_e B_m}p_{s_x}^2 \right)
\exp\left(-\frac{3}{2q_e B_m}p_{s_y}^2 \right) p_s E_s
\end{eqnarray}
The numerical evaluation  of  this  multi-dimensional  integral can be made
little bit easier (less time consuming job in computer) by assuming 
$E_3 \simeq \mu_s$ and $p_s \simeq p_{F_s}$ in
those parts of this integral where $\theta$, $\phi$  and  $E_3$  or
$p_3$  are  coupled together. The rate for reverse reaction is as
usual given by eqn.(19).\par 
To study chemical evolution of the
system, we have numerically solved the sets of kinetic  equations
given  by  eqns.(6)  and  (7).  As has already been mentioned, we
have considered three possible
scenarios: $(i)B_m=0$, $(ii) B_m^{(c)(e)} < B_m < B_m^{(c)(u,d)}$
and $(iii) B_m^{(c)(u,d)} <  B_m  <  B_m^{(c)(s)}$.  The  initial
conditions   are  $Y_u(t=0)=1.96$  and  $Y_d(t=0)=1.04$  for  the
fractional abundances of $u$ and $d$-quarks. The rates of various
processes  (eqns.(1)-(5))  are  also  evaluated  numerically   and
finally  the  charge  neutrality  and baryon number conservation
conditions (eqns.(8) and (9)) are used to relate $Y_s(t)$ and $Y_e(t)$, 
the fractional abundances for $s$-quark  and electron respectively with 
that of $u$ and $d$ quarks.  In fig.(1) we have plotted time evolution of
the fractional abundances $Y_i$ for $i=u,d,s,$  and  $e$  in  the
field  free  case. This figure shows that ultimately fractional abundances
for $u$, $d$, $s$-quarks and electron saturate to their equilibrium values. 
Which is the chemical equilibrium condition among the participants.  We have 
further noticed, that as the current mass $m_s$ for $s$-quark approaches 
the current mass of $u$ and $d$-quarks, the equilibrium values for fractional 
abundances of
all the three types of quarks ultimately saturates to a constant value
$\approx 1$ and the electron abundance $Y_e \approx 0$.  We have also
investigated the sensitivity of initial conditions. We have considered a
wide variation of proton fraction- from almost pure neutron matter
(proton fraction $x\approx 0.001$) to symmetric nuclear matter ($x=0.5$). The
qualitative nature of our results do not change with the change of
initial condition. The change in initial condition only causes a change
in saturation time scale. 
Now we would like to see whether such theoretical observation 
supports the speculation of Witten. We have verified whether the
system in $\beta$-equilibrium is energetically stable or not. To do that, 
we have followed the following steps. (i) Using these equilibrium
values for various components ($i=u,d,s,$ and $e$) we have  obtained  
the  equation of
states for the quark matter system with $B_m=0$, (ii) using the 
equation  of  state  we  have  solved  TOV equation and
obtained the mass of the quark core, which is  $\sim  0.4M_\odot$
for  the  central  density $4n_0$ and the corresponding radius is
$R=3.8$Km. To obtain the mass of quark core in the field free case, we
assume that the deconfinement transition is first order in nature. 
The corresponding critical density for a typical bag parameter value 
$B^{1/4}=160$MeV is $n_c\approx 2.1n_0$. It has been observed that 
the mass of the core increases in
presence of a magnetic field  where as the radius of the core decreases,
i.e. the core becomes more compact in presence of strong magnetic fields
\cite{sc7}. 
To investigate the stability of such a bulk quark matter, we have
calculated the energy per baryon using the relation
\begin{equation}
\frac{\epsilon}{n_B} =\frac{1}{n_B}\left ( \sum_{i=u,d,s,e} \epsilon_i +
\epsilon_{\rm{mag.}} \right )
\end{equation}
where $\epsilon_i$ is the internal energy density of the species $i$ and
$\epsilon_{\rm{mag.}}=B_m^2/(8\pi^2)$ is the magnetic energy density,
which is zero in the field free case. Using the saturation values for
abundances for various components, we have seen that $ < \epsilon_{QM}/n_B 
< M_n < \epsilon_{NM}/n_B$. Which indicates that the chemical evolution
of the system in field free case leads to a scenario which is
energetically more stable than nuclear matter and the binding energy of the
system is negative.  Therefore, the observation  is  in  agreement with the 
speculation of Witten. In  fig.(2) we  have  plotted the variation
of fractional abundances for various species with time for
$B_m=10^{14}$G, so that only electrons  are  affected.  In the
present scenario the deconfinement transition from hadronic matter to
quark matter can not be first order. We assume the transition is
metal-insulator type second order in nature. In this particular case, 
unlike a mixed phase of quark matter and hadronic matter in a first order
quark hadron phase transition, we have a sharp 
interface between nuclear matter crust and quark matter core. At the
sharp boundary all the thermodynamic variables change continuously, i.e.
$P_{NM}=P_{QM}$, $\epsilon_{NM}=\epsilon_{QM}$, $\mu_p=2\mu_u +\mu_d$
and $\mu_n=\mu_u +2\mu_d$. From these relations, we obtain the variation
of critical density with the strength of magnetic field. In fig.(3) we
have shown this variation. The corresponding bag parameter also changes
with the strength of magnetic field. The dashed curve in fig.(3) shows
the variation of $B^{1/4}$ with the strength of magnetic field. In both
the curves, up to $B_m= 10^3B_m^{(c)(e)}$ only electrons are affected
beyond which we have considered the quantum mechanical effect of strong
magnetic field on all the charged species except $s$-quark. In the
nuclear matter sector, we have considered a $\sigma-\omega$ exchange
type of mean field model \cite{sc5}. These curves show that both the
quantities remain almost constant (equal to their values for field free
case) up to $B_m=10^{17}$G, beyond which they become extremely large.
As it is obvious from fig.(2) that
in  this particular   case,  the  net production of $s$-quark  starts  after
$10^{-14}$sec, which is much earlier than the field free case, but
$Y_s$ goes to zero by $10^{-10}$sec and then it remains zero  for
ever.  It is also obvious from the figure that in this case $d$ quark 
abundance also goes to zero by the same time scale and in the chemical 
equilibrium condition it remains zero.
The system in chemical equilibrium is therefore mainly consists  of
$u$-quarks  and  electrons.  If  we translate this scenario
to the case of a neutron
star, this is equivalent to a proton rich  core.  As  a  consequence direct
URCA  process  will  play  the  main  role in the cooling process by neutrino
emission from this region, provided it is energetically stable \cite{sc8}. 
To study the stability of such abnormal system, we have
calculated internal energy per unit volume. Near nuclear  density
it is $\sim 4.39\times 10^{14}$MeV$^4$, which corresponds to free
energy  per baryon  $>>M_n$ and also $>>$ energy per baryon for hadronic 
matter.  The magnetic energy density is 
negligibly small compared to internal energy density of the system
unless the magnetic 
field is $>10^{19}$G. In this particular case, the overall energy of 
the system is
therefore equal to the kinetic energy of the system. This indicates that 
the system is therefore energetically
unstable if the magnetic field affects electrons present in the system. 
In fig.(4)
we have plotted the abundances for various species against time  
for  $B_m=5\times 10^{16}$G.
In  this  case  all  the  charged particles except $s$-quarks are
affected. We have seen that $s$-quark production is absolutely forbidden
in  this particular  scenario.  The system behaves like a neutron rich
(rich in $d$-quark) core. In  this  case  free  energy  per  unit
volume  is $\approx 4.9\times 10^{12}$MeV$^4$ near normal nuclear density, 
which again corresponds to
energy per baryon $>>M_n$ and also $>>$ energy per baryon for 
hadronic matter. Therefore, in this
case also, the system is energetically unstable. We have also noticed
that the presence of a small percentage of $\lambda$-hyperons in the 
core region before phase transition does not change the conclusions.
In figs.(5) and (6) we have plotted the abundances for the species with
time for the presence of $\approx 25\%$  $\lambda$-hyperon before
phase transition to quark matter for $B_m=0$ and $B_m=5\times 10^{16}$G
respectively.
We have noticed that for the field free case, the time evolution leads
to a stable configuration, where as for the presence of strong magnetic
field the system becomes energetically unstable in the chemical
equilibrium condition.
Therefore the presence of $\lambda$-hyperons in hadronic matter (which
is most likely if the density of hadronic matter exceeds a few times
normal nuclear density)
before deconfinement transition does not change our conclusion. 
Next we shall consider the effect of neutrino trapping on the stability
of quark matter core in presence of  strong magnetic fields in the next
section.
\section{Basic Formalism: With degenerate Neutrinos}
In this section we shall develop a formalism to investigate the
chemical evolution of the system considering the effect of neutrino
degeneracy on the weak reaction rates for the field free case and also
in presence of a strong magnetic field. Assuming the formation
of a neutron star immediately after supernova explosion, then at
the beginning the matter within the
remnant, particularly at the central region is extremely lepton rich
(neutrino trapped). Such an object is called
proto-neutron star. Normally it takes a few second to convert a
proto-neutron star to a neutron star by the emission of trapped
neutrinos. If the density of such matter exceeds $3-4n_0$, a
deconfinement transition to neutrino trapped quark matter can occur at
the core region. To study the chemical evolution of this neutrino
trapped quark matter, we have to evaluate the rates for the processes
\begin{equation}
d+\nu_e \leftrightarrow u+e^-
\end{equation}
\begin{equation}
s+\nu_e \leftrightarrow u+e^-
\end{equation}
and the direct and reverse reactions given by
\begin{equation}
u+d \leftrightarrow u+s
\end{equation}
In the case of trapped neutrinos in a degenerate quark matter we have to
solve the differential equations
\begin{equation}
\frac{dY_u}{dt}=-\frac{1}{n_B}[\Gamma_1^\prime +\Gamma_2^\prime]
\end{equation}
and
\begin{equation}
\frac{dY_d}{dt}=\frac{1}{n_B}[\Gamma_1^\prime -\Gamma_3^\prime]
\end{equation}
We also have to consider the total lepton number conservation, given by
\begin{equation}
Y_l =Y_e +Y_\nu \simeq 0.4
\end{equation}
The magic number $0.4$ is conventionally used in the numerical computation
of supernova explosion.
The baryon number conservation and charge neutrality conditions are also
used to obtain fractional abundances of all the species present in the
system. In eqns.(36) and (37), $\Gamma_i^\prime$'s are the
rates of the processes (33)-(35).
\subsection{Rate of Weak Processes: $B_m=0$}
In the case of degenerate neutrinos, the neutrino chemical potential
$\mu_\nu \ne 0$ and has to be evaluated from eqn.(38).
The rates for the processes given by eqns.(33) and (34) are given
by
\begin{equation}
\Gamma_1^\prime =\frac{G^2 \cos^2\theta_c}{4\pi^5}T^3
\xi_d^\prime(4\pi^2 +{\xi_d^\prime}^2) J(\mu_u ,\mu_e, \mu_d, \mu_\nu)
\end{equation}
and
\begin{eqnarray}
\Gamma_2^\prime &=& \frac{3G^2}{4\pi^5}\sin^2\theta_c T^3
(1-\exp(-\xi_s^\prime)) \int_{-\infty}^{(\mu_s -m_s)/T} dx \nonumber\\
& &\frac{(x+\xi_s^\prime)^2 +\pi^2}{(1+e^x)(1+e^-(x+\xi_s))} J(\mu_u-Tx,
\mu_e, \mu_s, \mu_\nu)
\end{eqnarray}
Finally the effective rate for the process (35) is given by
\begin{eqnarray}
\Gamma_3^\prime &=& \frac{9}{2\pi^5}G^2 \cos^2\theta_c \sin^2\theta_c
T^3 (1-\exp(-\xi^\prime_s)) \nonumber\\
& &\int_{-\infty}^{(\mu_s -m_s)/T} dx \frac{(x+\xi)^2
+\pi^2}{(1+e^x)(1+e^{-(x+\xi)})} J(\mu_u, \mu_s -Tx, \mu_u , \mu_d)
\end{eqnarray}
\subsection{Rate of Weak Processes: $B_m\neq 0$}
Now we shall consider the weak process in the degenerate quark matter in
presence of a strong magnetic field and also consider the effect of
neutrino neutrino trapping.
As before, we first assume that $B_m^{(c)(e)} <B_m < B_m^{(c)(u,d)}$, 
so that only
electrons are affected quantum mechanically by the magnetic field. In this
case the effective rate for the process (33) is given by
\begin{eqnarray}
\Gamma_1^\prime = \frac{G^2 \cos^2\theta_c}{\pi^3}(Tq_e B_m)
&&\sum_{\nu=0}^{\infty}\left(\frac{p_{F_u}}{p_{F_e}}\right)\mu_u \mu_e 
 \int_{x_u =0}^{\infty}dx_u \int_{0}^{\infty}dE_2
 \frac{1}{[\exp(-E_2^\prime /T-x_u)+1]} \nonumber \\ &&E_2^\prime
 f(x_u)[1+\exp(-(E_2-\mu_\nu)/T)]
 \end{eqnarray}
 where $E_2^\prime =E_2 -(\mu _u +\mu _e -\mu _d)$.\par
 To obtain rate for the process (34), we have to replace all the
 parameters of $d$-quark by the corresponding $s$-quark parameters and
 $\cos^2\theta_c$ has to be replaced by $\sin^2\theta_c$. In this case
 $E_2^\prime =E_2-(\mu_u +\mu_e -\mu_s)$.
 The rate for the process (35) is given by eqn.(41), which is the expression
 for field free case.\par
 When the magnetic field strength $B_m > B_m^{(c)(u,d)}$ but less than
 $B_m^{(c)(s)}$, all the charged species, except $s$-quarks are affected
 by the strong magnetic field. As has been mentioned before, compact
 analytical expressions for rates can only be obtained if we assume that
 only the zeroth Landau levels are populated  for all the
 magnetically affected charged particle components.  In this case the rate
 for the process (33) is given by
 \begin{eqnarray}
 \Gamma_1^\prime &=& \frac{3G^2}{2\pi^4}q_e B_m \left(\frac{q_u B_m q_d
 B_m}{17}\right)\cos^2\theta_c \frac{\mu_u \mu_e}{p_{F_u}p_{F_e}} T
 \nonumber\\
 & &\int_{x_e =-\infty}^{+\infty}\int_{E_2 =0}^{\infty}\int_{\theta
 =0}^{\pi}\int_{\phi =0}^{2\pi}\left(1-\frac{p_{F_d}p_{2_z}}{\mu_d
 E_2}\right)
\exp(-p_{2_x}^2/2q_e B_m) \exp(-9p_{2_y}^2/17q_e B_m) \nonumber\\
& & [1+\exp(-(E_2 -\mu_\nu)/T)]E_2^2 \frac{1}{[1+\exp(-x_e
-E_2^\prime/T)]}f(x_e)dx_e dE_2 \sin\theta d\theta d\phi
\end{eqnarray}
As before the rate for the process (34) can not be obtained simply by
replacing $d$-quark parameters by the corresponding $s$-quark values and
$\cos^2\theta_c$ by $\sin^2\theta_c$. In this case the rate is given by
\begin{equation}
\Gamma_2^\prime = \frac{5G^2}{16\pi^7}(q_e B_m q_u
B_m)\left(\frac{\mu_u}{p_{F_u}}\right)\left(\frac{\mu_e}{p_{F_e}}\right)
T^4 I \sin^2\theta_c
\end{equation}
\begin{equation}
I=\int_{-\infty}^{+\infty}\int_{-\infty}^{+\infty}\left(x_u +x_e
-\frac{\mu_s -\mu_u -\mu_e}{T}\right)^2 f(x_u)f(x_e) dx_u dx_e
\end{equation}
The rate for the process (35) is given by the combination of eqns.(31)
 and (19).\par
 In the degenerate neutrino case, we have plotted the time evolution of 
 fractional
 abundances for various species  in the filed free case in fig.(7). In
 this case the system is marginally stable, internal energy density is
 $\approx 6.25\times 10^9$MeV$^4$.  Since it takes a few second
 to convert such a system to conventional system of SQM with non-degenerate 
 neutrinos by neutrino emission, we may
 conclude that it is possible to have an energetically stable quark matter 
 core from the deconfinement transition at the central region of a compact 
 proto-neutron star. In the case when
 $B_m^{(c)(e)} < B_m < B_m^{(c)(u,d)}$, i.e., only electrons are
 affected, we have seen  that almost all the neutrinos are absorbed in
 the system momentarily and to keep lepton number constant, electron density
 increases by several orders of magnitude. As a consequence of charge
 neutrality condition, the $u$-quark density also increases by the same 
 orders of
 magnitude. But the conservation of baryon number of the system causes
 decrease in $d$-quark density almost by the same orders of magnitude.
 The situation is happened to be the most unphysical one. In this case 
 the binding energy calculation shows that the
 system is extremely unstable.
 In this case internal energy density is $\approx 10^{18}$MeV$^4$ near 
 normal nuclear density. In the case of strong magnetic
 field ($B_m^{(c)(u,d)} < B_m < B_m^{(c)(s)}$), when all the charged
 components except $s$-quark are affected, we have shown the time
 dependence of fractional abundances in fig.(8). In this case also
 $s$-quarks are never produced and the conclusion does not change if
 there are a few $\lambda$-hyperons at the central region of
 proto-neutron star before phase transition. We have also seen that
 electron density goes up by several orders of magnitude, as a
 consequence $u$-quark density also increases and the system behaves
 like a neutrino trapped proton core (if we translate it to
 proto-neutron star without quark core). The scenario is also
 unphysical in this case. The internal energy of the system is $\approx
 4.376\times 10^{17}$MeV$^4$. Therefore, we can conclude that in
 presence of a strong magnetic field, with trapped neutrinos, the formation of
 quark matter core by deconfinement transition at the central region of 
 a compact neutron star is also energetically forbidden.
 \section{Conclusions and Discussions}
We have studied the chemical evolution of quark matter core in presence of a 
strong magnetic field and compared our results with that of field free case. We
have noticed that in presence of a magnetic field of strength at least
greater than $B_m^{(c)(e)}$, the critical field strength to populate
Landau levels for electrons, quark matter core in  compact neutron star
or proto-neutron star is not possible. The quark matter core is
energetically unstable with respect to normal hadronic matter in presence 
of such strong magnetic field. On the
other hand if a charge neutral SQM in beta equilibrium is placed in an
external strong magnetic field  the system becomes energetically more
stable compared to field free case. We have also seen that the presence
of $\lambda$-hyperons at the core of a neutron star before phase
transition to deconfined quark matter does not alter the characteristic
of newborn quark matter core in presence of strong magnetic field.
We have also noticed that our conclusion is insensitive with the
numerical value of proton fraction.
Therefore, the final conclusion is that the quark matter core is not
at all possible  in a compact neutron star or proto-neutron star if the
magnetic field at the central region exceeds at least the critical value
for electron to populate Landau levels. In an earlier work we showed that
a first order QCD phase transition from nuclear matter to quark
matter is not possible in presence of a strong magnetic field \cite{sc0,sc1,sc2}. 
Even if a metal-insulator type second order phase transition is possible, the
stability criteria does not allow the formation of quark matter core in 
a compact neutron star or in a proto-neutron star. However, such a phase
transition is possible in the case of a very old neutron star by matter
accretion,  during slowing down of a pulser which may cause further 
contraction of neutron star, nucleation of SQM by some external sources or 
by neutrino sparking. Instead of a phase transition to quark matter it
is possible to have kaon condensed hadronic matter at the core region.
It would be interesting to see what is the effect of strong magnetic
field on such condensed state. It can also be a di-quark matter. It
would also be interesting to study the effect of strong magnetic field
on such matter.

\end{document}